\documentclass[oneside,twocolumn]{article}

\usepackage[sc]{mathpazo} 
\usepackage[T1]{fontenc} 
\usepackage{microtype} 

\usepackage[english]{babel} 

\usepackage[hmarginratio=1:1,top=20mm,columnsep=20pt]{geometry}
\usepackage[small,labelfont=bf]{caption} 
\usepackage{booktabs} 

\usepackage{lettrine} 

\usepackage{enumitem} 
\setlist[itemize]{noitemsep} 

\usepackage{abstract} 

\usepackage{titlesec} 
\titleformat{\section}[block]{\large\scshape\centering}{\thesection.}{1em}{} 
\titleformat{\subsection}[block]{\large}{\thesubsection.}{1em}{} 

\usepackage{titling} 

\usepackage[dvipdf]{graphicx}
\usepackage{tabularx,longtable,booktabs,array}
\newcolumntype{M}[1]{>{\centering\arraybackslash}m{#1}}
\usepackage{color}
\usepackage{amssymb}
\usepackage{subfigure}
\usepackage{natbib}

\usepackage{hyperref} 

\title{\bfseries A study of turbulence and interacting inertial modes \\ in a differentially-rotating spherical shell experiment} 
\author{%
\textsc{Michael Hoff} and \textsc{Uwe Harlander}
\\[1ex] 
\normalsize Department of Aerodynamics and Fluid Mechanics, \\ Brandenburg \normalsize University of Technology (BTU) Cottbus - Senftenberg, \\ \normalsize Siemens-Halske-Ring 14, 03046 Cottbus, Germany \\ 
\normalsize \href{mailto:uwe.harlander@b-tu.de}{uwe.harlander@b-tu.de} 
\and 
\textsc{Santiago Andr\`{e}s Triana} \\[1ex] 
\normalsize Royal Observatory of Belgium, \\ 
\normalsize Ringlaan 3, 1180 Brussels, Belgium \\
\normalsize \href{mailto:santiago.triana@oma.be}{santiago.triana@oma.de} 
}
\date{\today} 
\renewcommand{\maketitlehookd}{%
\begin{abstract}
\noindent We present a study of inertial modes in a differentially rotating spherical shell (spherical Couette flow) experiment with a radius ratio of $\eta = 1/3$. Inertial modes are Coriolis-restored linear wave modes which often arise in rapidly rotating fluids. Recent experimental work has shown that inertial modes exist in a spherical Couette flow for $\Omega_{i}<\Omega_{o}$, where $\Omega_i$ and $\Omega_o$ is the inner and outer sphere rotation rate. A finite number of particular inertial modes has previously been found. By scanning the Rossby number from $-2.5 < Ro = (\Omega_{i}-\Omega_{o})/\Omega_{o} < 0$ at two fixed $\Omega_{o}$, we report the existence of similar inertial modes. However, the behavior of the flow described here differs much from previous spherical Couette experiments. We show that the kinetic energy of the dominant inertial mode dramatically increases with decreasing Rossby number that eventually leads to a wave-breaking and an increase of small-scale structures at a critical Rossby number. Such a transition in a spherical Couette flow has not been described before. The critical Rossby number scales with the Ekman number as0 $E^{1/5}$. Additionally, the increase of small-scale features beyond the transition transfers energy to a massively enhanced mean flow around the tangent cylinder. In this context, we discuss an interaction between the dominant inertial modes with a geostrophic Rossby mode exciting secondary modes whose frequencies match the triadic resonance condition. 
\end{abstract}
}

\begin{document}

\maketitle


\section{Introduction}\label{sec:1}

The fluid between two concentric spheres differentially rotating around a common axis, nominally the spherical Couette flow, is relevant for geophysical and astrophysical objects like planetary interiors. This is because many planetary bodies consist of a solid inner and a liquid outer core which do not rotate at a constant angular velocity \citep{spohn_2007}. The Earth's core, for example, undergoes a slight differential rotation \citep{song_richards_1996,buffett_1997}. It is helpful to quantify the interaction between the core rotation and the fluid's interior in order to understand e.g. angular-momentum transport, tidal heating, fluid mixing or the generation of magnetic fields.

One common feature in rapidly rotating systems, like planetary spherical shells, are inertial waves which are Coriolis-restored internal oscillations \citep{greenspan_1968}. Already in the 19th century, Poincar\'{e} derived the governing equations for inertial waves. Without boundaries, plane inertial waves exist for any frequency in the interval $0 \le \omega \le 2\Omega$, where $\omega$ is the inertial wave frequency and $\Omega$ the angular velocity of the rotating fluid, and the frequency alone defines the angle between the wave vector and the rotation axis \citep{greenspan_1968}. Inertial waves contained in arbitrary containers (e.g. cylindrical or spherical annulus) are called \emph{inertial modes}. An implicit analytical solution for inertial modes in a spheroid has been found by Bryan \cite{bryan_1889}. In a rotating fluid sphere, analytical solutions for inertial modes in the inviscid limit (zero viscosity) have been found by Kudlick \cite{kudlick_1966} and, recently, Zhang \textit{et al.} \cite{zhang_et_al_2001} found an explicit solution for the inertial-mode velocity field. In contrast, for a spherical shell, as similar as full sphere and shell geometry may appear, no analytical solution for inertial modes exists because the hyperbolic Poincar\'{e} equation does not comply with the common no-slip boundary conditions \citep{zhang_et_al_2001}. Much effort has been taken to find numerical and analytical alternatives in solving the hyperbolic equation, e.g. ray tracing \citep{tilgner_1999,maas_2001,harlander_maas_2007,harlander_maas_2007b,rieutord_valdettaro_2010,koch_et_al_2013,rabitti_maas_2013,hoff_et_al_2016}. Viscosity regularizes the singular solutions typical for spherical shells. It is required for the excitation of inertial modes since it transforms the boundary-layer singularities at the critical latitudes into detached shear layers \citep{kerswell_1995}. These critical-latitude singularities, where the wave characteristics are tangential to the inner sphere's boundary, play an important role in periodically forced flows, e.g. librational \citep{calkins_et_al_2010,seelig_2014,hoff_et_al_2016} or tidal forcing \citep{rieutord_valdettaro_2010,sauret_et_al_2014}. 

The above-mentioned results of the last few decades are mainly of analytical and numerical nature. Therefore, laboratory experiments with fluid-filled spheres and spherical shells give a decisive advantage in validating previous theoretical considerations or to find new aspects of the flow. Inertial modes can be excited by different mechanisms. One of the pioneering experimental work was done by Adridge and Toomre \cite{aldridge_toomre_1969}. For a libration-driven flow in a full sphere, they found a set of inertial modes with azimuthal and axial wavenumbers $m$ and $l$ that match with the solutions from the corresponding boundary-value problem. Later, Aldridge \cite{aldridge_1972} confirmed that some of the inertial oscillations have their counterparts in thick spherical shells. Based on seismological data, Aldridge and Lumb \cite{aldridge_lumb_1987} found that large-scale inertial modes (small azimuthal wavenumber) may exist in the Earth's liquid outer core, however, this was contested soon after. In the following years, a number of laboratory experiments in spheres and spherical shells with different excitation mechanisms have been done, e.g. inner sphere libration \citep[e.g.][]{koch_et_al_2013,hoff_et_al_2016}, outer sphere libration \citep[e.g.][]{noir_et_al_2009}, tidal deformation \citep[e.g.][]{sauret_et_al_2014}, precession \citep[e.g.][]{triana_et_al_2012} or a free oscillating inner sphere \citep{kozlov_et_al_2015}. 

Most recently, it has been found that inertial modes can be excited also in differentially forced spherical-gap flows \citep{kelley_et_al_2007,kelley_et_al_2010,zimmerman_et_al_2011,triana_2011} although no periodic external forcing is applied. One would expect that inertial modes exist for all frequencies up to $2\Omega$. In contrast, Kelley \textit{et al.} \cite{kelley_et_al_2007} showed first that only a small finite number of particular inertial modes are excited by differential rotation. Surprisingly, the structure of these modes is similar to their full-sphere counterparts \citep{zhang_et_al_2001}. It was speculated that the selection of these particular modes might be related to \emph{over-reflection} which causes a mode amplification by extracting energy from the shear flow induced by the differential rotation. Kelley \textit{et al.} \cite{kelley_et_al_2010} extended this work and found particular modes that are excited by \emph{critical-layer} resonance that is related to over-reflection \citep{rieutord_et_al_2012}. First numerical simulations, related to the experiments by Kelley \textit{et al.} \cite{kelley_et_al_2007}, have been done by Matsui \textit{et al.} \cite{matsui_et_al_2011}. These authors were able to estimate the azimuthal velocity inside the tangent cylinder and found that the velocity outside is almost in solid-body rotation with the outer shell. Since most of the fluid volume ($\sim 85\%$) in a spherical shell with $\eta = r_i/r_o = 1/3$ is located outside the tangent cylinder, they concluded that this might be one reason for the structural similarity between the full-sphere modes and the spherical-shell modes, as long as the modes possess a weak amplitude inside the tangent cylinder. Triana \cite{triana_2011} investigated inertial modes in a much bigger spherical shell ($r_{o}=1.46$m) operating at Ekman numbers of $E = \nu / (\Omega_{o}r_{o}^2)\ge2.5\times 10^{-8}$ that is in the range of the asymptotic regime ($E\rightarrow0$), relevant in the planetary and stellar context. In accordance with this, Rieutord \textit{et al.} \cite{rieutord_et_al_2012} did numerical experiments, but without differential rotation (again arguing that most of the fluid volume is in solid-body rotation with the outer shell), which revealed resonant peaks at frequencies that are roughly in agreement with the experiments. They advanced a hypothesis based on \emph{critical layers} within the Stewartson layer, where the phase speed of the inertial modes matches the angular velocity of the flow. These layers occur at a certain maximum Rossby number ($Ro_{max}<0$) above which inertial modes with a particular azimuthal wavenumber $m$ cannot be excited. 

In contrast to flows driven by periodic forcing, the experiments in a differentially rotating spherical shell did not reveal internal shear layers. Nonetheless, first numerical simulations with a real, but very small, differential-rotation forcing \citep{baruteau_rieutord_2013} approved the existence of internal shear layers what eventually leads to short-period attractors by multiple reflections. 

The present experimental work builds on previous experimental studies on differentially rotating spherical shells but focus on the stability of the inertial modes that has not been experimentally studied before. Once the inertial modes got excited, our experiments reveal a strong amplification of the most dominant inertial mode, leading to a transition into small-scale disorder of the flow structure due to a secondary instability which has not been observed in any of the previous studies. The Rossby number where this transition occurs, we refer to as \emph{critical Rossby number} $Ro_c$, scales approximately with $E^{1/5}$. In the present paper, we largely focus on this transition and just briefly describe features of previous studies we can confirm by our experiments.

The remainder of this paper is organized as follows: We start by a brief description of the experimental apparatus, the measurement technique and the data processing in section~\ref{sec:2}. Section~\ref{sec:3} shows experimental results. We discuss the spectrograms (amplitude spectrum as a function of frequency and Rossby number) and connect them to prominent flow features (section~\ref{sec:3a}). Some dominant wave modes in the counter-rotation regime interact with each other leading to secondary peaks which satisfy the triadic-resonance conditions (section~\ref{sec:3b}). The zonal mean flow and its dependency on the Rossby number will be described in section~\ref{sec:3c}. In section~\ref{sec:3d}, we focus on the scaling and the kinetic energy distribution of the transition where a large-amplitude wave breaks and gives rise to weak small-scale disorder. Finally, a discussion and conclusion of the new flow features follows in section~\ref{sec:4}.

\section{Experimental Set-up and Data Processing}\label{sec:2}

\subsection{Experimental Set-up}\label{sec:2a}

\begin{figure}
	\includegraphics[width=8cm]{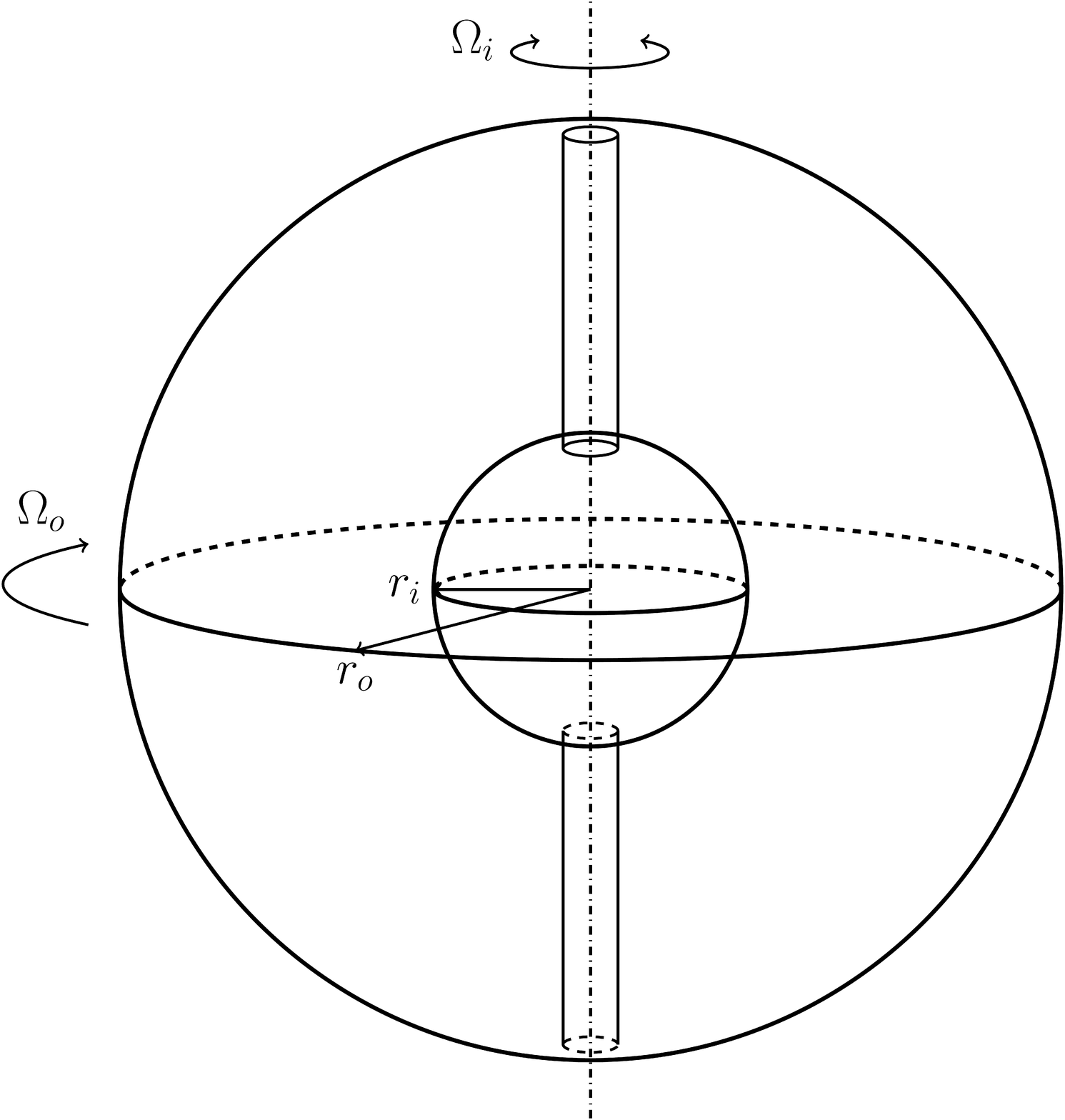}
	\caption{Sketch of the spherical shell setup. The outer sphere is rotating clockwise with constant speed (as seen from top) around a vertical axis. The inner sphere rotation is variable.}
	\label{fig:1}
\end{figure}
The experimental apparatus (sketched in figure~\ref{fig:1}) consists of two independently rotating concentric spheres with inner radius $r_i=(40\pm 0.05)\,\mathrm{mm}$, outer radius $r_o=(120\pm 0.05)\,\mathrm{mm}$ and a corresponding gap width of $d = (80\pm 0.1)\,\mathrm{mm}$. From this follows a radius ratio of $\eta=1/3$ that is similar to that of the Earth's inner and outer core $\eta_{core}=0.35$ \citep{spohn_2007}. The inner sphere is made of black anodized aluminum suspended on a shaft of 14$\,$mm diameter while the outer sphere is made of acrylic glass with full optical access except at the equator where the two hemispheres are connected. We used a silicone oil of viscosity $\nu_{kin}=0.65\,\mathrm{mm^2s^{-1}}$ ($\pm 10\%$ tolerance) as the working fluid in the gap. To avoid optical distortions and keep the surrounding temperature uniform, the shell is immersed into a cubic tank of de-ionized water (refraction indices: $n_{oil}=1.375$ and $n_{water}=1.337$ for 532$\,$nm). The outer and inner sphere rotation is denoted by $\Omega_o$ and $\Omega_i$, respectively. By using $(\Omega_i-\Omega_o)\,r_o$ as velocity scale and $r_o$ as length scale, a characteristic Rossby number of $Ro = (\Omega_i-\Omega_o)/\Omega_o$ for steady differential rotation can be defined. The Ekman number $E = \nu_{kin} / (\Omega_o r_o^2)$ measures the ratio of viscosity and Coriolis forces. With our apparatus, we can achieve values of $E\ge 6.8\cdot 10^{-6}$. 

The flow in the horizontal plane has been studied quantitatively with \emph{Particle Image Velocimetry} (PIV). Spherical hollow glass spheres have been used as tracer particles. Two GoPro Hero 4 cameras, enabling wireless high-resolution recordings, observed the motion of the particles \emph{in the frame at rest with the outer shell}. With the present set-up, the flow in approximately 40\% of the horizontal plane between the spheres can be observed. 

\subsection{Data and Data Processing}\label{sec:2b}

We recorded two particular experimental ramps for two fixed outer sphere rotation rates at $\Omega_o\approx(32,64)\mathrm{rpm}$. The inner sphere rotation was in the range $-45 \le \Omega_i \le +24$ and $-90 \le \Omega_i \le +48$ for the respective $\Omega_o$ so that we cover $-2.5 \le Ro \le -0.2$ in the Rossby-number space. 
\begin{figure*}
	\includegraphics[width=8.6cm]{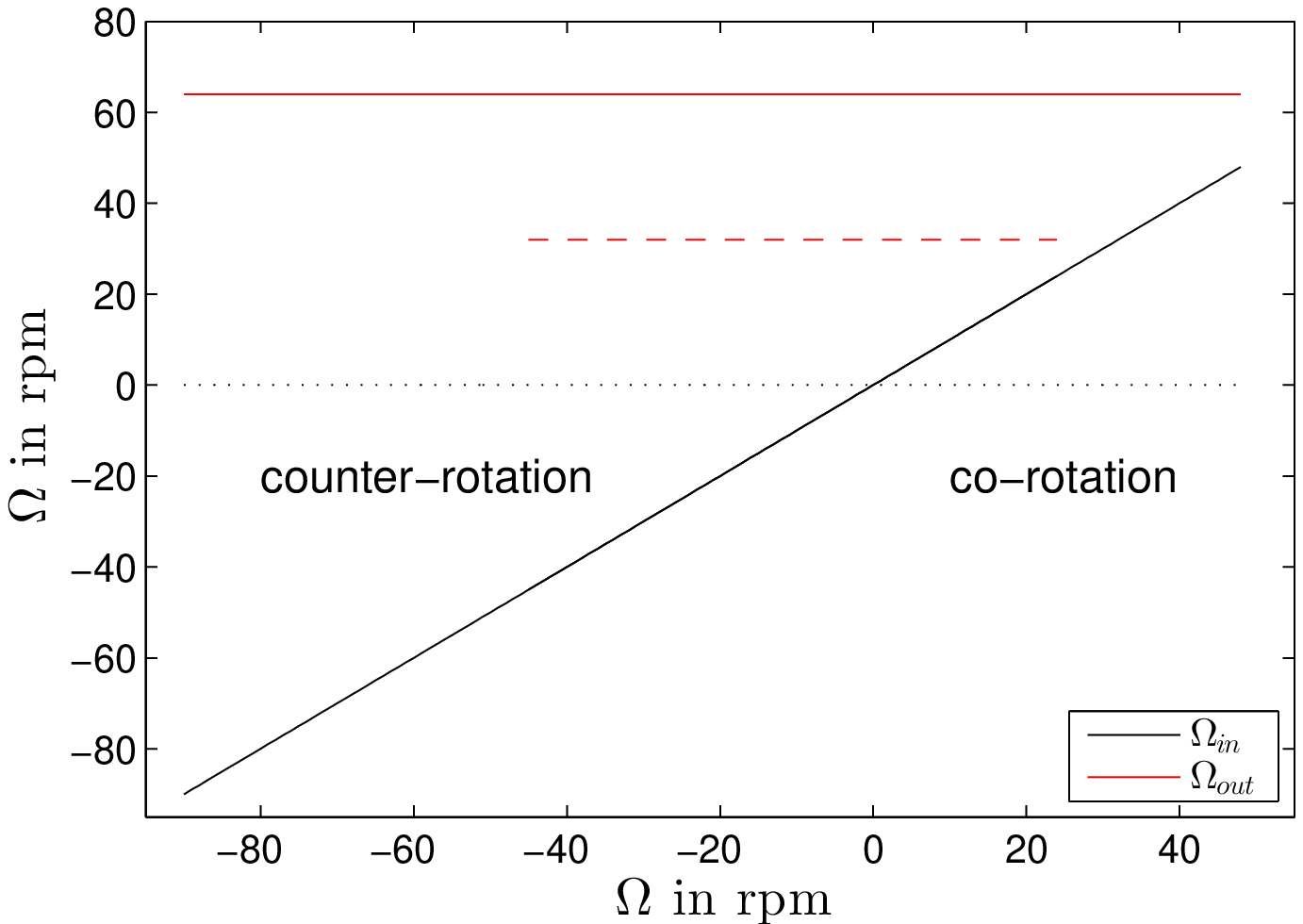}
	\includegraphics[width=8.6cm]{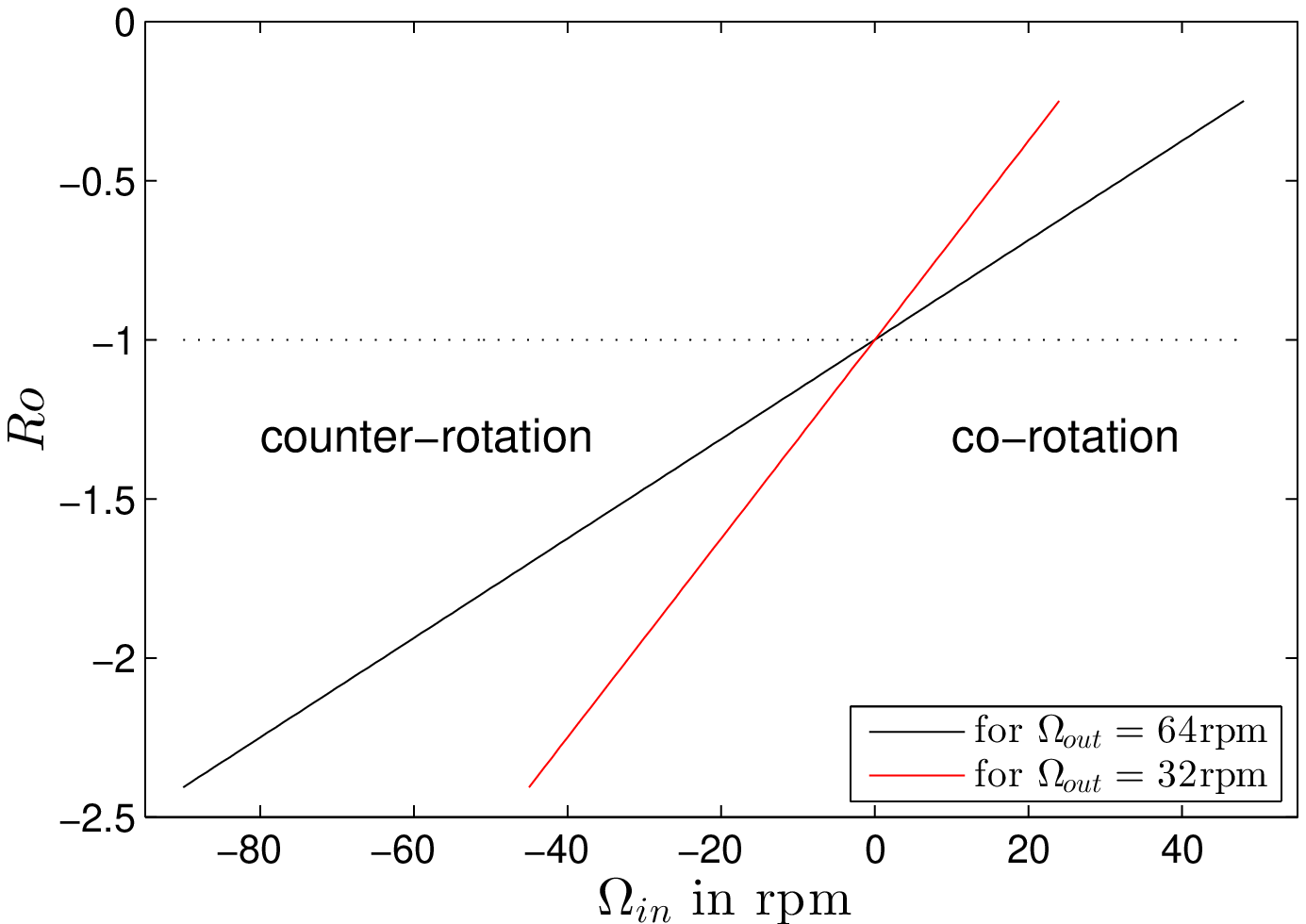}
	\caption{Graphical illustration of the experimental parameters; outer and inner sphere rotation as a function of $\Omega$ (upper) and the Rossby number as a function of the inner sphere rotation (lower). The experimental ramps have been performed from left to right along the lines. }
	\label{fig:2}
\end{figure*}
Figure~\ref{fig:2} schematically illustrates the performed parameter ramps. Each ramp started in the counter-rotation regime at $Ro=-2.5$ where we let the apparatus run for about 15 minutes to avoid transient spin-up recirculation. Then we increased the inner sphere rotation by increments of $\Delta Ro \approx 1/30$ (or $\Delta \Omega_i \approx 1(2)\mathrm{rpm}$ for $\Omega_o \approx 32(64)\mathrm{rpm}$). At each particular step, we waited 5 minutes to ensure an equilibrium state and recorded the flow for 15 minutes in the horizontal laser plane at height 4cm above the equator, i.e. tangential to the north pole of the inner sphere. Note that the spin-up time from rest ($\sim E^{-1/2}\Omega_o^{-1}$) is in the order of 1 to 2 minutes \citep{greenspan_1968}. Due to a fast discharging of the camera batteries, we were not able to record the full ramps without interruptions. After at least 6 steps in the ramp, the engines needed to be stopped and the batteries of the cameras recharged. After that, the ramp has been started again at $Ro=-2.5$, followed by slowly increasing $Ro$ until the breaking point.

The movies of the horizontal plane have been converted into gray scale images and analyzed by using the Matlab toolbox \emph{matPIV v.1.6.1} \citep{sveen_2004}. For the present purpose, a spatial resolution of 1920$\times$1080 pixel was sufficient to obtain reliable velocity fields. For $\Omega_o\approx 32(64)\mathrm{rpm}$ a frame rate of 15(30) fps has been used in the co-rotation regime ($-1.0 \le Ro \le -0.2$) and 30(60) fps in the counter-rotation regime ($-2.5 \le Ro \le -1.0$). We used three interrogation steps from $128\times 128$ to $64\times 64$ to a final window size of $32 \times 32$ with an overlap of 0.5. A signal-to-noise filter, a peak height filter and a global filter that removes vectors significantly larger or smaller than the majority of the vectors, have further been applied. A Fourier analysis was applied to the velocity components $(u,v)$ to detect the dominant frequencies in the flow. For these frequencies, the corresponding flow patterns have been studied by a harmonic analysis. The harmonic analysis is a signal-demodulation technique in which the user specifies wave frequencies to be analyzed and applies least-square techniques to find the unknown amplitudes and phases of the waves (see e.g. \cite{emery_thomson_2001}, chapter 5.5).

\section{Results}\label{sec:3}

\subsection{The Spectrograms}\label{sec:3a}

Before discussing the spectrograms, we should comment on the inertial mode notation. Each mode can be written as a spherical harmonic $Y_l^m$, where the degree $l$ and the order $m$ are symmetry numbers \cite{greenspan_1968}. Here, $m$ corresponds to the azimuthal and $l$ to the axial wavnumber. Further, each mode has a unique frequency $\hat{\omega}=\omega/\Omega_o$. This notation $(l,m,\hat{\omega})$ has been used in any of the previous studies \citep{zhang_et_al_2001,kelley_et_al_2007,kelley_et_al_2010,matsui_et_al_2011,triana_2011,rieutord_et_al_2012}.

With this definition we will next discuss results from a Fourier analysis of the two experimental ramps.
\begin{figure*}
	\centering
	\includegraphics[width=16cm]{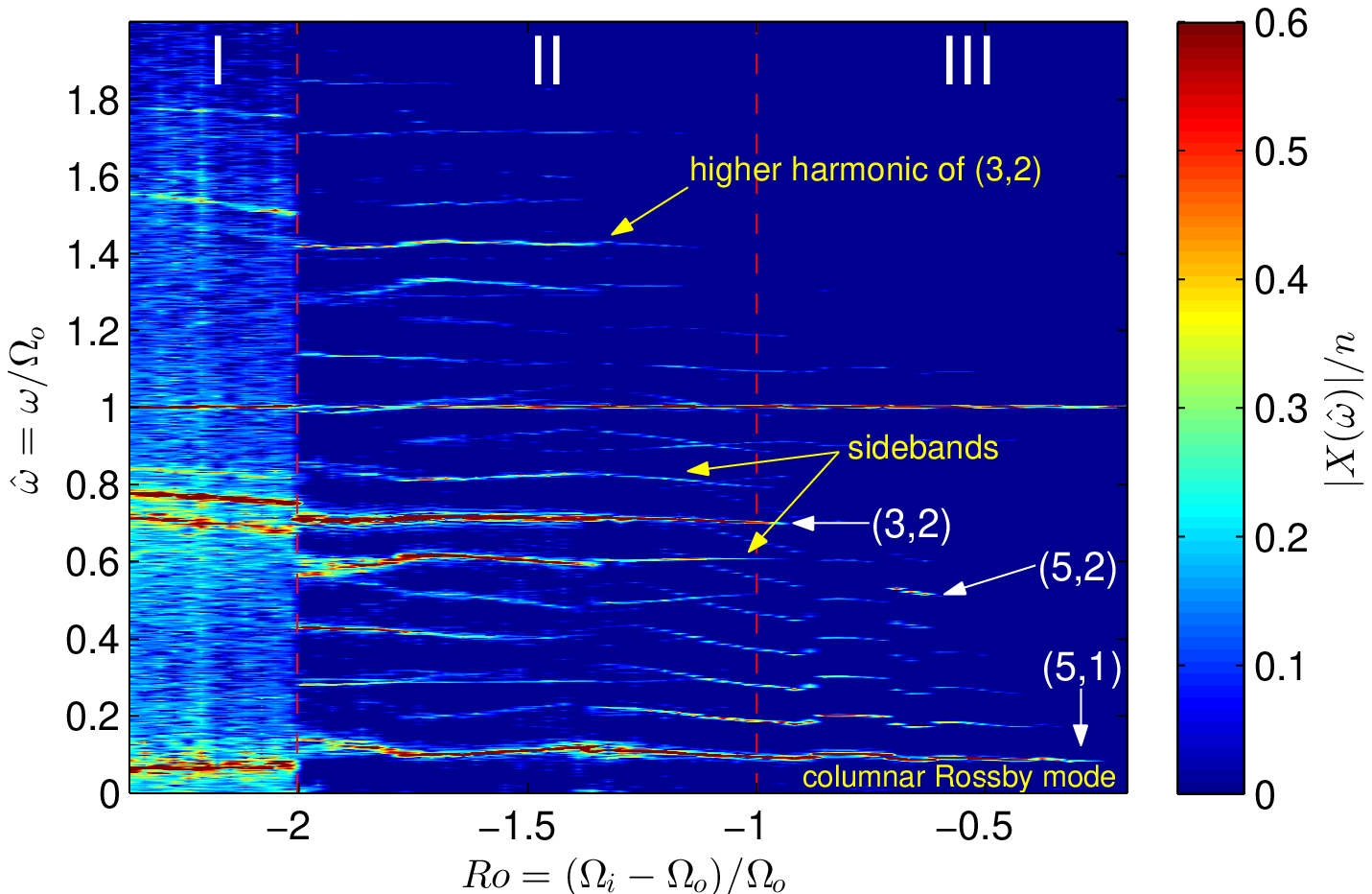}
	\caption{Azimuthal-velocity spectrogram taken from a ramp where each inner sphere rotation was kept constant for 20 minutes (including 5 minutes spin-up time). The outer sphere rotation is $\Omega_o\approx 32\mathrm{rpm}$ and the corresponding Ekman number $E=1.35\cdot 10^{-5}$. At each grid point in a radial cross section, a Fourier transform of the azimuthal velocity has been computed where the velocities are averaged over 5 neighboring grid points. Each column of the spectrogram shows the average over all obtained amplitude spectra $|X(\hat{\omega})|/n$. The labels (I), (II) and (III) mark the three regimes discussed in the text.}
	\label{fig:3}
	\includegraphics[width=16cm]{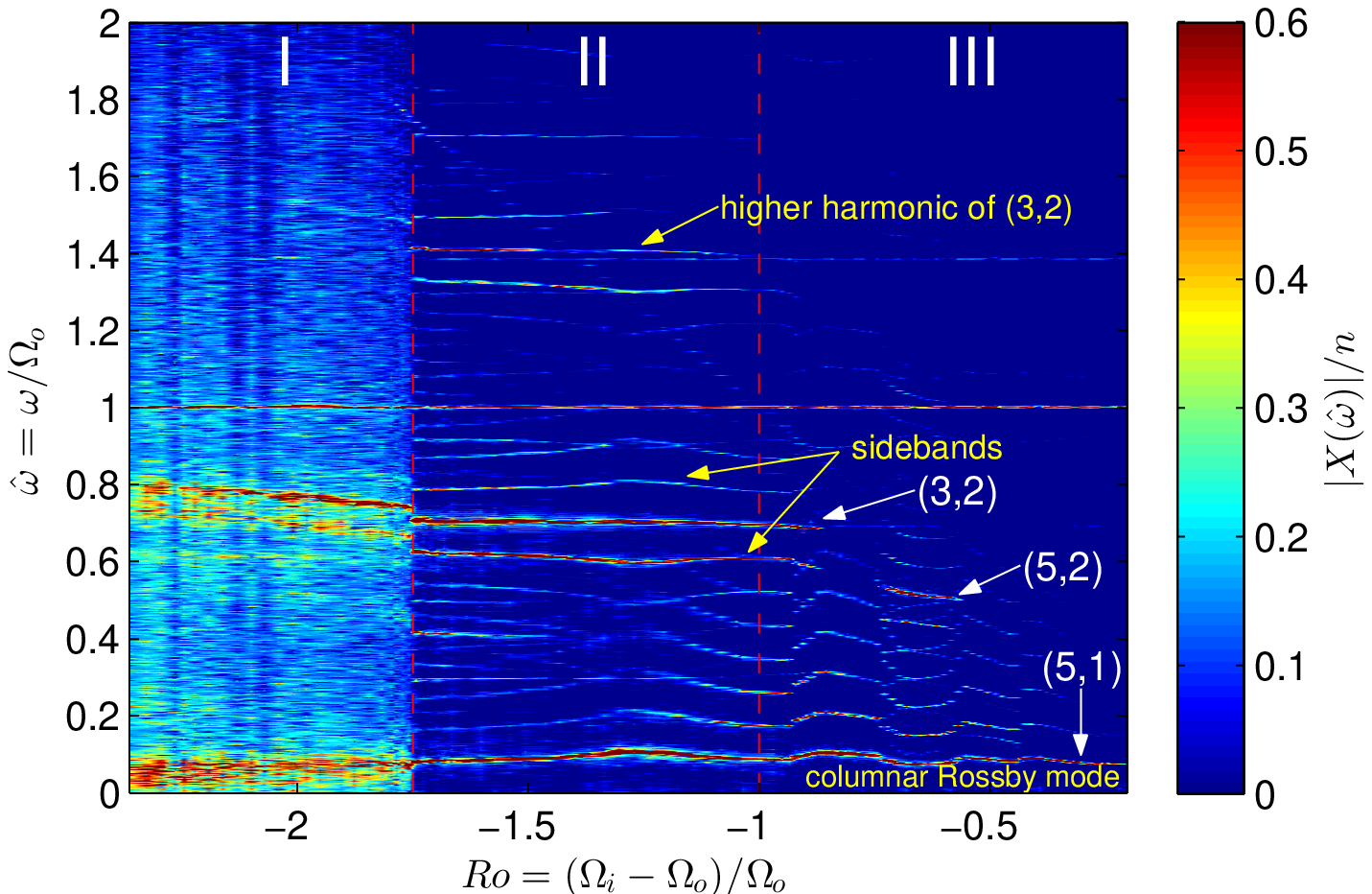}
	\caption{The same as Figure~\ref{fig:3} but for $\Omega_o\approx 64\mathrm{rpm}$ and $E=6.76\cdot 10^{-6}$.}
	\label{fig:4}
\end{figure*}
Figure~\ref{fig:3} and \ref{fig:4} show spectrograms of the azimuthal velocity for $\Omega_o\approx 32\mathrm{rpm}, E=1.35\cdot 10^{-5}$ and $\Omega_o\approx 64\mathrm{rpm}, E=6.76\cdot 10^{-6}$, respectively. The data have been taken \emph{in the frame at rest with the outer shell}. The spectrograms show the single-sided amplitude spectra $|X(\hat{\omega})|/n$, where $n$ is the number of time steps, as a function of the dimensionless inertial wave frequency $\hat{\omega}=\omega/\Omega_o$ in the range $0\le\hat{\omega}\le 2$ versus the Rossby number $Ro$. Therefore, each column of $Ro$ represents a radius-averaged amplitude spectrum of 15 minute time series of the azimuthal velocity where the velocity has been smoothed by averaging over an area of $-0.6\mathrm{cm} \le x \le +0.6\mathrm{cm}$ (about 10 grid points).

Roughly speaking, both spectrograms can be separated into three parts labeled as (I), (II) and (III) on top of figures~\ref{fig:3} and \ref{fig:4}. The first one is the strong counter-rotation regime characterized by a higher turbulence level, broader, and also weaker and blurred signals. We found that this regime is dominated by small-scale fluctuations, especially around the tangent cylinder explaining the higher turbulence level. Therefore, we coin this regime as \emph{weakly-turbulent regime} (I).

Mainly three dominant frequencies stand out in this regime. The lowermost peak could be identified as the low-frequency columnar Rossby mode $(l=5,m=1,\hat{\omega}_1\approx 0.09)$ which is the first Stewartson-layer instability setting in after solid-body rotation \cite{hollerbach_2003,wicht_2014}. This mode is persistent over the entire Rossby number range considered. The highest of the dominant frequencies could be identified as the $(l=3,m=2,\hat{\omega}_0\approx 0.75)$ inertial mode which shows a slight linear drift $\hat{\omega} = -0.12\,Ro + 0.54$ (in agreement with \cite{rieutord_et_al_2012}). The peak in between has a wavenumber of $m=1$ and satisfies the condition $\hat{\omega}_0 = \hat{\omega}_1 + \hat{\omega}_2$ and $m_0 = m_1 + m_2$ indicating a three-wave coupling due to triadic resonance between the modes. Taking a closer look to the spectrograms in figures~\ref{fig:3} and \ref{fig:4}, this triad is found to be prominent over the entire counter-rotation region ($Ro\lesssim -1$). We examine this mode interaction in more detail in section~\ref{sec:3b}. 

The noisy regime ends very abruptly at a Rossby number that we call critical Rossby number $Ro_c$, indicated by the leftmost vertical red line in figures~\ref{fig:3} and \ref{fig:4}. For $\Omega_o \approx 32(64)\,\mathrm{rpm}$, a critical Rossby number of $Ro_c \approx -2.0 (-1.73)$ can be found. To our knowledge, such a clear transition from weak small-scale turbulence to a more organized structure is observed for the first time in a spherical Couette flow. Therefore, the main body of this paper focuses on this transition.

After passing the critical Rossby number we enter regime (II) for which the turbulence level is lower compared to the weakly-turbulent regime (I). After the transition, several prominent peaks pop out in the entire inertial-wave frequency range. We found that this regime is dominated by mostly well-organized discrete large-scale structures leading to a higher signal amplitude. Therefore, we coin this regime as \emph{strong-inertial-mode regime} (II). The three dominant modes detected in regime (I) are preserved but we found a slight change of their modal structure to larger scales after passing $Ro_c$. The peak at $\hat{\omega}=1.4$ is a higher harmonic of the (3,2) mode with wavenumber $m=4$ \citep{koch_et_al_2013,hoff_et_al_2016}. In contrast, in the experiments by \cite{triana_2011,rieutord_et_al_2012} operating with much smaller Ekman numbers, peaks at frequencies $\hat{\omega}>1.0$ have not been observed since for low Ekman numbers, the higher harmonics do in general not survive the transition to regime (I) (see also figure~\ref{fig:4}).  

\begin{figure}
	\centering
	\includegraphics[width=8.6cm]{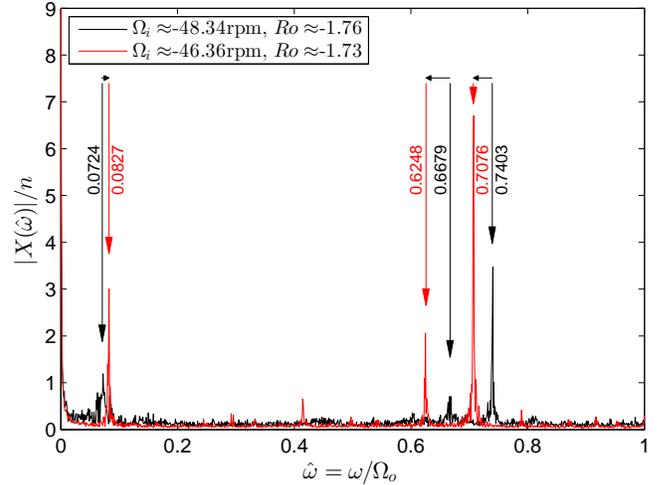}
	\caption{Two amplitude spectra $|X(\hat{\omega})|/n$ extracted from figure~\ref{fig:4} for $Ro = -1.76$ (black) in the weakly-turbulent regime and $Ro=-1.73$ (red) in the strong-inertial-mode regime. The outer sphere rotation rate is $\Omega_o\approx 64\mathrm{rpm}$ and $E=6.76\cdot 10^{-6}$. The numbers represent the frequency of the peaks and the horizontal arrows mark the frequency shift after transition from regime (I) to (II).}
	\label{fig:5}
\end{figure}
Remarkable is further the frequency jump when passing the critical Rossby number. To highlight this feature, figure~\ref{fig:5} shows the amplitude spectra $|X(\hat{\omega})|/n$, extracted from the spectrogram for $\Omega_o\approx 64\mathrm{rpm}$ (figure~\ref{fig:4}) at two different Rossby numbers, $Ro \approx -1.76$ in the weakly-turbulent regime (black curve), and $Ro \approx -1.73$ in the strong-inertial-mode regime (red curve). The vertical arrows indicate the three dominant peaks in both regimes and the horizontal arrows indicate the frequency jump that occurs after passing from the weakly-turbulent regime (I) into the strong-inertial-mode regime (II). Note that the change in $Ro$ corresponds to a $\Delta \Omega_i$ of just 2rpm. After the transition from (I) to (II), the peaks become sharper, the signal amplitude $|X(\hat{\omega})|/n$ is increased by a factor of about 2 and the turbulence level is decreased. Note that a similar behavior can be observed for $\Omega_i \approx 32\mathrm{rpm}$ but at a smaller critical Rossby number $Ro_c \approx -2.0$. The frequency jump arises from a Doppler shift due to an abruptly changing shear flow around the tangent cylinder after passing the critical Rossby number. We will examine this in more detail in section~\ref{sec:3c}.

The third regime in figures~\ref{fig:3} and \ref{fig:4} is the \emph{co-rotation regime} (III) in the range $-1 < Ro < -0.2$ where the inner sphere is rotating slower than the outer shell but in the same direction (rightmost vertical red line). Most of the modes that were dominant in the strong-inertial-mode regime become weak and eventually disappear in the co-rotation regime. Beside $\hat{\omega}=1$ resulting from technical shortcomings, the only permanently persisting strong peak is the columnar Rossby mode (5,1). Furthermore, a number of higher harmonics $k\,\hat{\omega},\ k=1,2,3,...$ of this mode with decreasing signal amplitude for higher $k$ emerge in this regime. Another rather dominant peak appears at $\hat{\omega}= 0.52\pm 0.015$ within $-0.73\le Ro \le -0.6$ which could be identified as the $(5,2)$ inertial mode. Within its Rossby-number range, the $\hat{\omega}-Ro$ dependency of this mode fits to $\hat{\omega} = -0.20\,Ro + 0.39$ for 60rpm (figure~\ref{fig:4}) which again is in agreement with \cite{rieutord_et_al_2012}. Close to solid-body rotation ($Ro > -0.2$), all frequency peaks vanish due to a shut down of any instability in this weak shear flow.

We finally note, in agreement with \cite{rieutord_et_al_2012}, that all identified modes propagate retrograde, i.e. against the rotation of the outer shell, all $(l-m)$ odd modes are non-axisymmetric and antisymmetric with respect to the equator, and the columnar Rossby mode with $(l-m)=4$ is symmetric with respect to the equator.
All these modes are annotated in figures~\ref{fig:3} and \ref{fig:4}. The frequency range, Rossby number range, drift speed and the type of the instability \citep{wicht_2014} for the determined modes are summarized in Table~\ref{tab:2}. In addition, the modes we found correspond also with the modes from previous experimental \citep{kelley_et_al_2007,kelley_et_al_2010,zimmerman_et_al_2011,matsui_et_al_2011,triana_2011,rieutord_et_al_2012} and numerical results \citep{wicht_2014}.

\begin{table*}
	\centering
	\caption{Uniquely identified inertial modes in regime (II) and (III). Identification by comparison of the frequency with the model from \cite{zhang_et_al_2001}, \cite{wicht_2014}, and the experimental works from \cite{kelley_et_al_2007,kelley_et_al_2010,zimmerman_et_al_2011,matsui_et_al_2011,triana_2011}, and \cite{rieutord_et_al_2012}. Additionally, the spatial patterns have qualitatively been compared with the modes found by \cite{zhang_et_al_2001}. $Ro_{max}$ marks the onset of the particular mode in the inviscid limit taken from \cite{rieutord_et_al_2012}. The abbreviations mean: ES - Equatorial Symmetric, EA - Equatorial Antisymmetric, SL - Stewartson-layer instability, and geo - geostrophic (columnar pattern, $z$-invariant).}
	\begin{tabular}{M{1cm} M{2cm} M{2.5cm} M{2.5cm} M{2.5cm} M{2.5cm} p{2cm}}
		\hline\multicolumn{7}{l}{\textbf{(a) $\mathbf{\Omega_o=31.87rpm}$, $\mathbf{E = 1.35\cdot 10^{-5}}$}} \\ \hline
		$(l,m)$ & $\omega_{ana}/\Omega_o$ & $\omega_{meas}/\Omega_o$ & $(\omega_{meas}/\Omega_o)/m$ & $|Ro|$ range & $|Ro_{max}|$ \cite{rieutord_et_al_2012} & instability \\ \hline
		(3,2) & 0.6667 & 0.715 - 0.698 & 0.357 - 0.349 & 1.989 - 0.945 & 0.70 & SL, EA \\
		(5,1) & -0.0682 & 0.140 - 0.081 & 0.140 - 0.081 & 1.989 - 0.252 &  & SL, ES (geo) \\
		(5,2) & 0.4669 & 0.529 - 0.510 & 0.265 - 0.26 & 0.693 - 0.600   & 0.50 & SL, EA \\
		\hline\multicolumn{7}{l}{\textbf{(b) $\mathbf{\Omega_o=63.74rpm}$, $\mathbf{E = 6.76\cdot 10^{-6}}$}} \\ \hline
		$(l,m)$ & $\omega_{ana}/\Omega_o$ & $\omega_{meas}/\Omega_o$ & $(\omega_{meas}/\Omega_o)/m$ & $|Ro|$ range & $|Ro_{max}|$ \cite{rieutord_et_al_2012} & instability \\ \hline
		(3,2) & 0.6667 & 0.708 - 0.688 & 0.353 - 0.345 & 1.727 - 0.888 & 0.70 & SL, EA \\ 
		(5,1) & -0.0682 & 0.108 - 0.074 & 0.108 - 0.074 & 1.727 - open & & SL, ES (geo) \\
		(5,2) & 0.4669 & 0.536 - 0.501 & 0.268 - 0.254 & 0.729 - 0.569 & 0.50 & SL, EA \\ 
	\end{tabular} 
	\label{tab:2}
\end{table*}

\subsection{Triad interactions}\label{sec:3b}

Triad interactions occur for internal gravity waves in stratified fluids \citep{bourget_et_al_2013,bourget_et_al_2014} and inertial waves in precessing cavities \citep{albrecht_et_al_2015,lin_et_al_2015} or even in rigidly rotating fluids \citep{bordes_et_al_2012}.

As mentioned before, we detected a dominant interaction between the (3,2) mode and the columnar Rossby mode (5,1) over a large Rossby number range generating a secondary peak satisfying $\hat{\omega}_0 = \hat{\omega}_1 + \hat{\omega}_2$ and $m_0 = m_1 + m_2$. In the following, we refer to these three dominant inertial modes in the spectrograms as \emph{dominant triad}. The (3,2) mode with frequency $\hat{\omega}_0$ dominates the triad and is nearly independent of $Ro$ in regime (II). In contrast, a change of frequency $\hat{\omega}_1$ of the Rossby mode (5,1) leads to a change  in $\hat{\omega}_2$ of the secondary peak, or vice versa. Another secondary, but weaker peak in regime (II) of figures~\ref{fig:3} and \ref{fig:4} can be found around $\hat{\omega}_2' \approx 0.79$ slightly above the frequency of the (3,2) mode which also satisfies the triadic condition $\hat{\omega}_0 = \hat{\omega}_2' - \hat{\omega}_1$. 

This implies that the interaction between an inertial mode and the columnar Rossby mode gives rise to two sidebands (labeled in figures~\ref{fig:3} and \ref{fig:4}) of this inertial mode, forming triads. Having a closer look at the spectrograms, such sidebands could also be found for all other $(l-m)$ odd inertial modes, i.e. the (5,2) and the higher harmonic of the (3,2) mode. The sideband with the higher (lower) frequency always exhibits an azimuthal wavenumber of $m_{sideband}=m_{mode}\pm1$. Apparently, the sideband with the lower frequency is more prominent than the one with the higher frequency. The reason is that the smaller-scale patterns (higher azimuthal wavenumber) have a larger damping and dissipate faster. In addition, especially for the dominant (3,2) mode, also very weak secondary sidebands can be detected below and above the respective primary sidebands. 

To quantify the wave interactions in regime (II) and (III) we use a bispectral analysis \citep{nikias_raghuveer_1987}. With the help of the HOSA Matlab toolbox, we computed the bicoherence defined by
\begin{equation}
B(\hat{\omega}_1,\hat{\omega}_2) = \frac{X(\hat{\omega}_1)X(\hat{\omega}_2)X^\ast(\hat{\omega}_1+\hat{\omega}_2)}{\sqrt{|X(\hat{\omega}_1)|^2|X(\hat{\omega}_2)|^2|X^\ast(\hat{\omega}_1+\hat{\omega}_2)|^2}},
\end{equation}
where $X(\hat{\omega})$ is the Fourier transform and $^\ast$ denotes the complex conjugate, for each grid point along a radial cross-section (the same as for the spectrograms) over five time windows of about $270\,\mathrm{s}$ with an overlap of 50$\,$\%. All obtained bicoherence spectra were then averaged. With such a normalization the bicoherence gives a statistical measure of the quadratic phase coupling \citep{nikias_raghuveer_1987,brouzet_et_al_2016} with 'zero' for random phases and no coupling, and 'one' for perfect coupling.
\begin{figure}
	\centering
	\includegraphics[width=8.5cm]{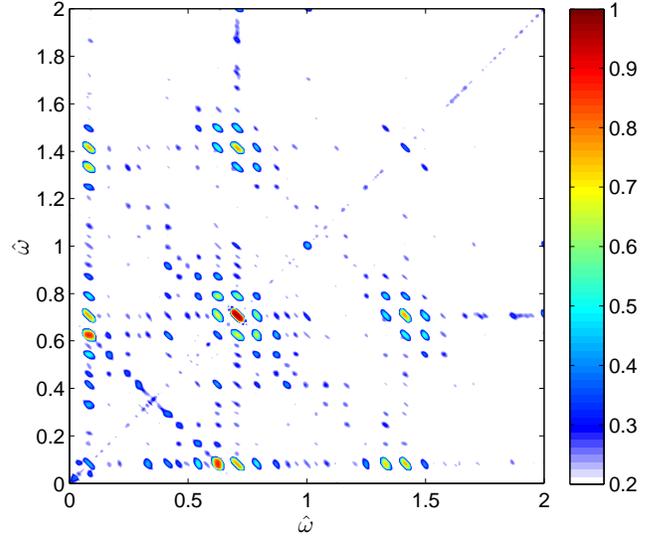}
	\caption{Bicoherence calculated for $E=6.76\cdot 10^{-6}$ and $Ro\approx-1.73$ in regime (II) for the same spatial region as the spectrogram in figure~\ref{fig:4}.}
	\label{fig:6}
\end{figure}
Figure~\ref{fig:6} shows the Bicoherence spectrum calculated for $E=6.76\cdot 10^{-6}$ and $Ro\approx-1.73$ in regime (II) slightly before the transition to regime (I). The most dominant peak is at $\hat{\omega}=0.7$ (self-correlation of the (3,2) mode). Possible triplets related to this mode can be found along the line with slope -1 between the points (0,0.7) and (0.7,0). From this, we directly see the dominant triad (0.70,0.61,0.09) as the red spots with correlation $\gtrsim 80\%$. Further, secondary waves are acting as primary waves for higher-order triadic resonances. Based on this, we can also detect numerous triadic interactions between different modes when choosing other $\hat{\omega}_0$ as a primary wave. Similar to \cite{brouzet_et_al_2016}, this result is a cascade of triadic interactions transferring energy from large-scale features with large amplitude to many discrete inertial waves with smaller amplitude. Therefore, triadic resonances play an important role in our spherical Couette flow, especially in regime (II) and (III).

As mentioned, the bicoherence spectrum is taken slightly before the transition to regime (I) and hence shows the last, and most complex state (high number of subharmonic instabilities) before the flow passes into regime (I) for which only the dominant triad found in regime (II) survives the transition (not shown). Instead, the ``continuous part" or the background noise increases suddenly which resembles a cascade to small-scale wave turbulence (further discussions in section~\ref{sec:3d}).

\subsection{The Zonal Mean Flow}\label{sec:3c}

\begin{figure*}
	\includegraphics[width=8.6cm]{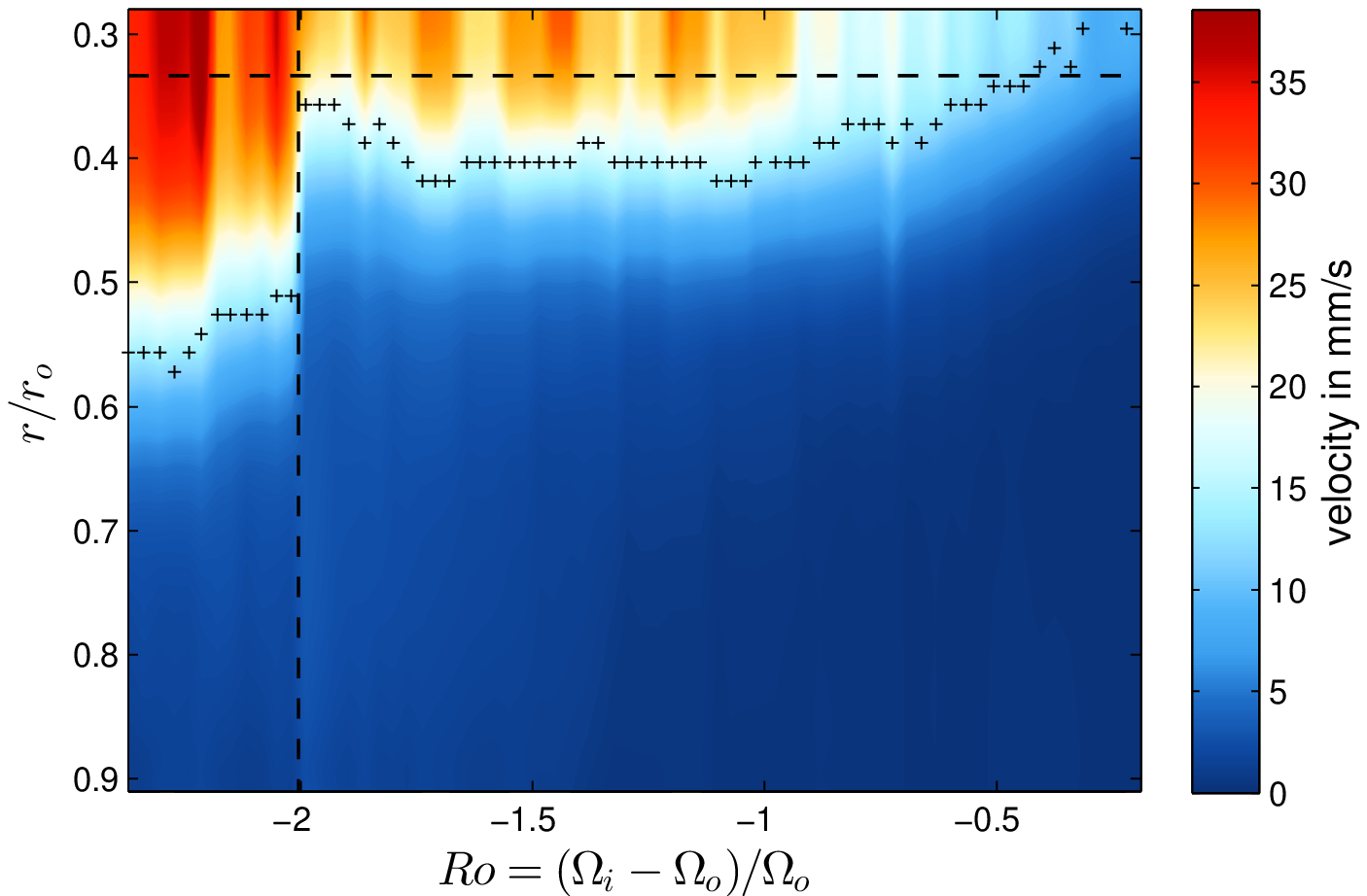}
	\includegraphics[width=8.6cm]{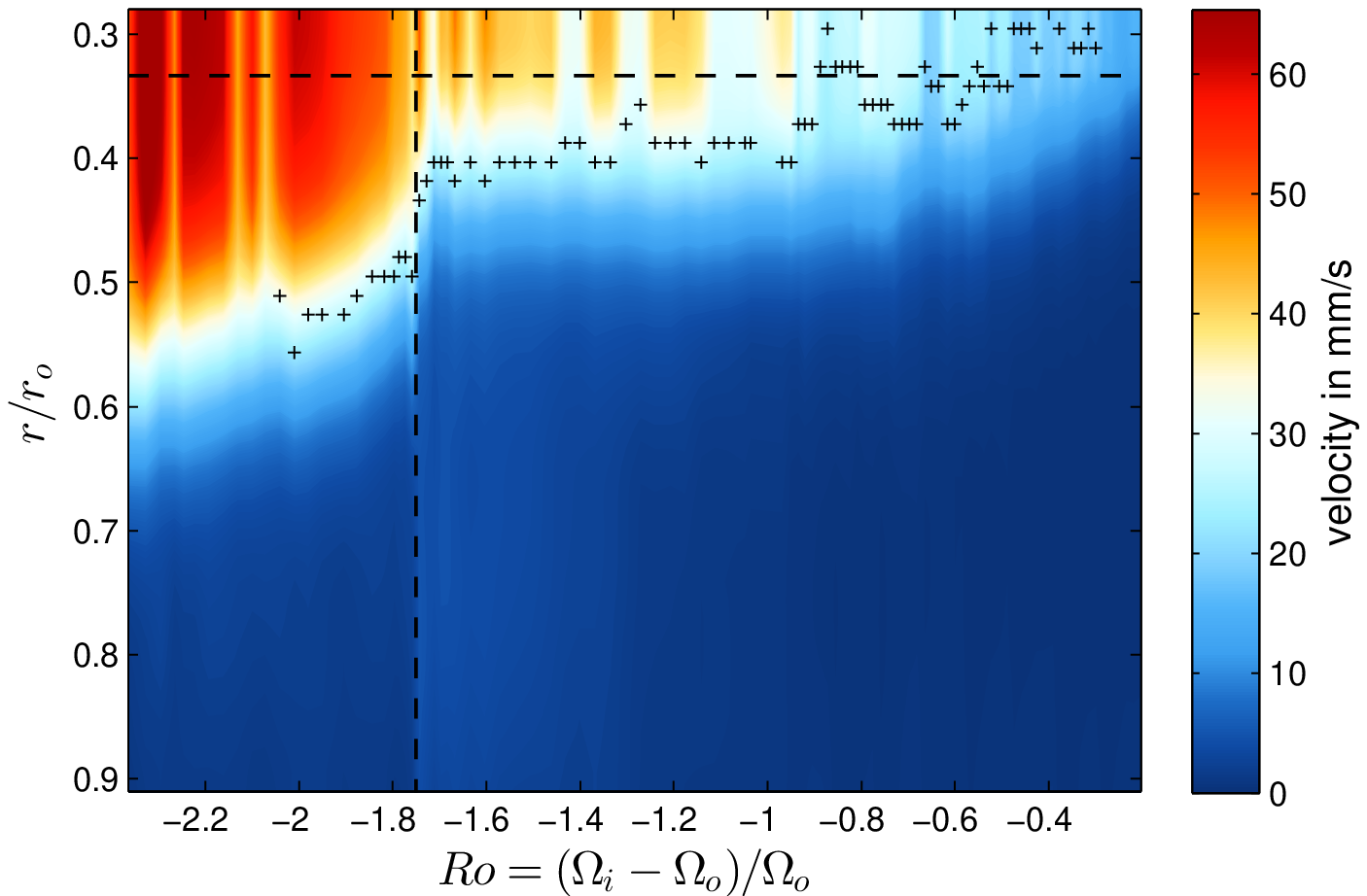}
	\caption{Temporal and azimuthal average of the azimuthal velocity $(\overline{v_{\phi}}^{t,\phi})$ in mm/s taken from a ramp where each inner sphere rotation was kept constant for 20 minutes (including 5 minutes spin-up time). Left: $\Omega_o\approx 32$, $E=1.35\cdot 10^{-5}$. Right: $\Omega_o\approx 64$, $E=6.76\cdot 10^{-6}$. The symbols mark the radial position of critical layers of the (5,1) mode.}
	\label{fig:7}
\end{figure*}  
In the following we examine how the transition at $Ro_c$ impacts the zonal mean flow. In contrast to other opaque spherical Couette flow experiments \cite{kelley_et_al_2007,kelley_et_al_2010,triana_2011,rieutord_et_al_2012}, we have optical access to almost 40\% of the horizontal plane in the spherical gap and can measure properly resolved radial profiles of the azimuthal velocity. Figure~\ref{fig:7} shows the azimuthally- and time-averaged azimuthal velocity taken from the two ramps for $\Omega_o\approx 32$rpm, $E=1.35\cdot 10^{-5}$ (left) and $\Omega_o\approx 64$rpm, $E=6.76\cdot 10^{-6}$ (right). Each column represents a time average over 15 minutes at a particular inner sphere rotation while the outer sphere's rotation was kept fix. Due to the slower rotating inner sphere ($Ro<0$), the mean flow in the frame at rest with the outer shell is always retrograde. The symbols mark the radial position of critical layers \cite{rieutord_et_al_2012} of the (5,1) mode. They will be discussed in section~\ref{sec:4}.

Basically, there are significant fluctuations in velocity magnitude. So far, we do not have a clear understanding of this feature. It might result from wave-wave or wave-mean flow interactions or is a spurious effect that comes from restarting the experiment for each $Ro$. 

Roughly speaking, when starting from regime (I), there is a continuous decrease of the mean velocity with increasing $Ro$ resulting from the decreasing differential rotation. However, due to the presence of inertial waves, there are decisive differences to profiles that would have been expected without any wave activity. First, the highest experimentally obtained velocities in both cases (Figure~\ref{fig:7} upper left $\sim 40\mathrm{mm\ s^{-1}}$ and Figure~\ref{fig:7} upper right $\sim 65\mathrm{mm\ s^{-1}}$) are located in the weakly turbulent regime (I) where the flow is supercritical. At the two critical Rossby numbers $Ro_c \approx -2.0$ (left) and $Ro_c \approx -1.73$ (right), marked by the vertical dashed lines, the velocity and the radial extension of the strong flow zone drops suddenly by 30-50\% when crossing $Ro_c$. Averaged over $10\ \Delta Ro$ intervals left- and right-hand side of $Ro_c$, we obtain $\Delta \langle v_{\phi} \rangle = \langle v_{\phi}(I) \rangle - \langle v_{\phi}(II) \rangle = (8.9\pm 4.2)\,\mathrm{mm/s}$ and $\Delta \langle v_{\phi} \rangle = (13.9\pm 6.0)\,\mathrm{mm/s}$ for $E=1.35\cdot 10^{-5}$ and $E=6.76\cdot 10^{-6}$, respectively, where $\langle \cdot \rangle$ denotes the average over time, radius and azimuth and the errors are the standard deviation over the $10\ \Delta Ro$ intervals. The errors are large, however, this is not surprising since the mean flow in regime (II) shows high fluctuations and in regime (I) the mean flow is gradually increasing which causes a large standard deviation.

The abrupt change in the mean flow at $Ro_c$ affects the frequency of the wave modes. As mentioned in section~\ref{sec:3a}, we suggest that the frequencies at $Ro_c$ change due a Doppler shift \cite{pedlosky_1987} according to
\begin{equation}
\hat{\omega}(I) \Omega_o = \hat{\omega}(II) \Omega_o + \Delta \overline{U}\,m/r_o,
\end{equation}
where $\hat{\omega}(I)$ and $\hat{\omega}(II)$ are the frequencies in regime (I) and (II), $r_o$ is the length scale for $m$ and $\Delta \overline{U}\,m/r_o$ is the Doppler shift. Taking the frequency of the (3,2) mode averaged over $10\ \Delta Ro$ on either side of $Ro_c$ (see figure~\ref{fig:3}, \ref{fig:4} or \ref{fig:5}), we obtain $\Delta \overline{U} = (9.9\pm 1.3)\,\mathrm{mm/s}$ ($\Delta \overline{U} = (16.0\pm 1.9)\,\mathrm{mm/s}$) for $E=1.35\cdot 10^{-5}$ ($E=6.76\cdot 10^{-6}$). Despite the large errors, the match between $\Delta \overline{U}$ and $\Delta\langle v_{\phi} \rangle$ is fairly well, supporting the assumption of a Doppler shift due to the abruptly changing mean flow. Note that the match for the (5,1) mode is similar, but not for the sideband at $\hat{\omega}=0.62$ since this peak changes only according the triadic resonance condition of the primary wave modes (3,2) and (5,1) discussed in section~\ref{sec:3b}.

Furthermore, after passing $Ro_c$ to regime (II), the mean flow remains almost constant until $Ro\approx-1$, i.e. shortly after the inner sphere changed the direction of rotation. We will see later that this happens due to a significant growth of the (3,2) mode in this regime. 

In the co-rotation regime ($-1.0 \le Ro \le -0.2$) the retrograde mean flow is monotonically decreasing due to decreasing shear and a shut down of the wave activity. 

\subsection{Critical Rossby number scaling, energy, and wave turbulence}\label{sec:3d}

From section~\ref{sec:3a}, we know that the critical Rossby number increases for smaller Ekman numbers. In order to find a possible scaling of the critical Rossby number with respect to $E$, we performed additional experiments for outer sphere rotation rates with $\Omega_o \approx (55,50,45,40,35)\mathrm{rpm}$ so that we finally have 7 data points for $Ro_c$. The ramps have been performed again by starting at $Ro = -2.5$ followed by a smooth increase of the inner sphere rotation rate with increments of $\Delta \Omega_i = 1\mathrm{rpm}$. Inspecting the values of $Ro_c$ from Figures~\ref{fig:3} and \ref{fig:4}, we expect that $Ro_c$ lies somewhere in between [-2.0,-1.73] for the outer rotation rates given above. As before, we recorded data of the flow in the horizontal plane 4cm above the equator for each value of $\Omega_i$. The transition to the strong inertial mode regime, and hence $Ro_c$, can visually be detected using the following criteria: (i) the (3,2) mode becomes dominant (see section~\ref{sec:3a}), (ii) the zonal mean flow strongly decelerates and the width of the Stewartson layer decreases abruptly (see section~\ref{sec:3c}). 
\begin{figure}
	\centering
	\includegraphics[width=9.5cm]{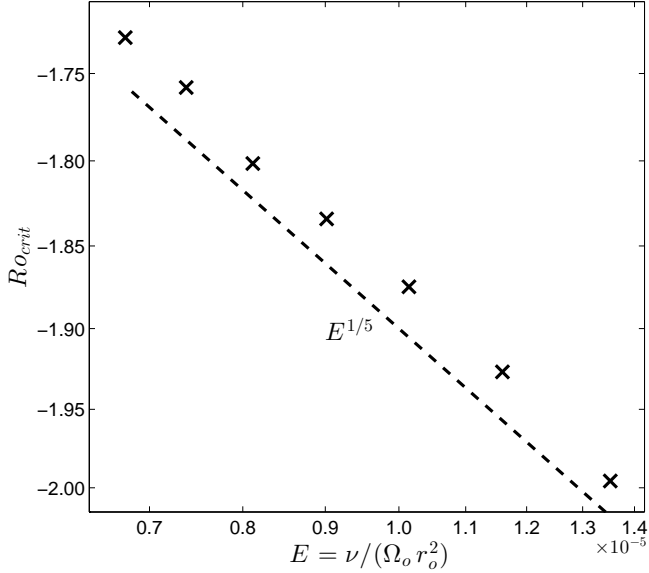}
	\caption{Critical Rossby number $Ro_c$ (crosses) where the weak turbulence transitions into a strong-inertial-mode regime as a function of the Ekman number. The crosses at highest $E$ and lowest $E$ correspond to the observations in the figures \ref{fig:3} and \ref{fig:4}, respectively. The dashed line shows the Ekman-number scaling $E^{1/5}$ of the critical Rossby number.}
	\label{fig:8}
\end{figure}
In Figure~\ref{fig:8}, we plot the critical Rossby number $Ro_c$ (crosses) versus the Ekman number $E=\nu/(\Omega_o r_o^2)$ in log-log-scale. The dashed line shows the $E^{1/5}$ dependency. Obviously, $Ro_c$ scales approximately with $E^{1/5}$. Following this $E^{1/5}$ power law, the critical Rossby number at $E= 10^{-8}$, for which the other spherical-gap experiments have been performed \citep{kelley_et_al_2007,kelley_et_al_2010,zimmerman_et_al_2011,matsui_et_al_2011,triana_2011,rieutord_et_al_2012}, is about $Ro_c \approx -0.47$. This implies that most of these experiments were running in the weakly turbulent regime. However, this regime has not been described for the previous experiments. Note that for numerically accessible Ekman numbers of about $E\ge 10^{-4}$ \citep[e.g.][]{hollerbach_2003,wicht_2014}, the power law would imply a $Ro_c = -2.95$, which infers a strong counter-rotation.

\begin{figure}
	\centering
	\includegraphics[width=8.6cm]{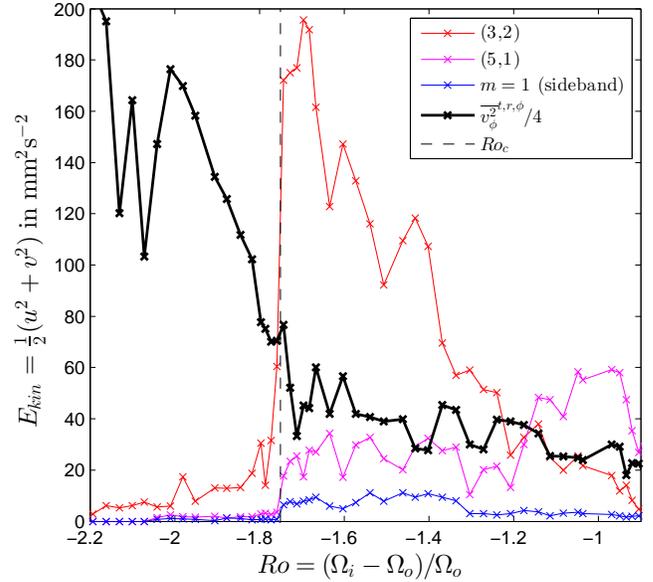}
	\caption{Maximum kinetic energy of the dominant triad (3,2), (5,1) and lower sideband as a function of the Rossby number for $\Omega_o \approx 64\mathrm{rpm}$ and $E = 6.76\cdot 10^{-6}$. The vertical dashed black line marks the critical Rossby number $Ro_c$. The solid black curve shows the energy of azimuthally and radially averaged mean flow $\langle v_{\phi}^2 \rangle$.}
	\label{fig:9}
\end{figure} 
We shall now examine the transition from a kinetic energy point of view. Figure~\ref{fig:9} shows the maximum kinetic energy of the dominant triad (red, magenta and blue line) for the $\Omega_o\approx 64\mathrm{rpm}$ ramp as a function of the Rossby number. The data have been filtered harmonically by each of the respective frequency $\hat{\omega}$. The thick black line shows the kinetic energy of the radially and azimuthally averaged mean flow scaled by a factor of 4 for a better display. 

It is remarkable that on the right hand side of the critical line ($Ro>Ro_c$), the kinetic energy of the (3,2) mode (red line) is gradually increasing with decreasing $Ro$ until reaching $Ro_c$. In contrast, the energy of the columnar (5,1) mode has its maximum at $Ro\approx -1$ where the (3,2) is weak. The sideband of the dominant triad shows an almost constant and low energy spectrum compared to the other modes. Moreover, the mean flow (thick black line) remains almost constant during the growth phase of the (3,2) mode until $Ro_c$.

At $Ro_c$, the energy of the dominant modes, especially of the (3,2) mode, drops dramatically by a factor of about 10 and then further decreases with decreasing $Ro$. In contrast, the mean flow shows a significant enhancement for decreasing $Ro<Ro_c$. Consequently, there is a clear ``knee" in the energy profile of the mean flow around $Ro_c$ deviating from an originally expected linearly increasing mean flow with decreasing Rossby number. This implies that for $Ro>Ro_c$ the (3,2) mode might draw energy from the shear flow and hence suppresses the growth of the shear flow around the tangent cylinder. For $Ro<Ro_c$, after the modes ``break'', giving rise to small-scale turbulence, the transfer of energy between the small scales and the zonal mean flow becomes more efficient, leading to an enhanced shear flow around the tangent cylinder.

Finally, we give some details on the turbulence that occurs after the transition to regime (I). 
\begin{figure}
	\centering
	\includegraphics[width=8cm]{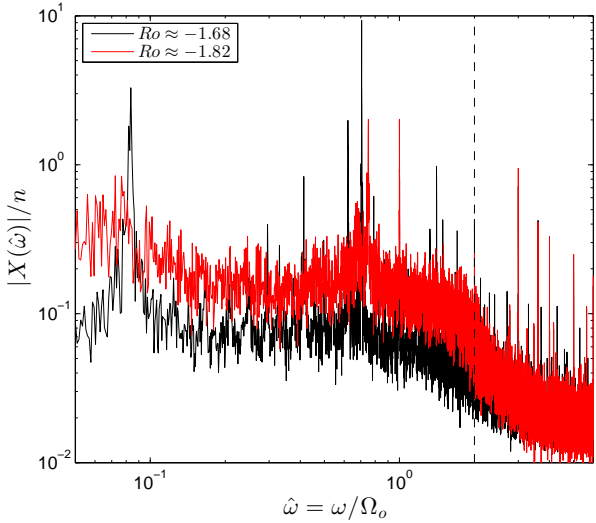}
	\caption{Two amplitude spectra $|X(\hat{\omega})|/n$ in log-log scale extracted from figure~\ref{fig:4} for $Ro = -1.82$ (red) in the weakly-turbulent regime (I) and $Ro=-1.68$ (black) in the strong-inertial-mode regime (II). The critical Rossby number is at $Ro_c = -1.73$. The outer sphere rotation rate is $\Omega_o\approx 64\mathrm{rpm}$ and $E=6.76\cdot 10^{-6}$. The dashed vertical line marks $\hat{\omega}=2$, beyond which the inertial waves become evanescent.}
	\label{fig:10}
\end{figure}
Figure~\ref{fig:10} shows two amplitude spectra in log-log scale for $\Omega_o=64\,\mathrm{rpm}$ and $Ro \approx -1.68$ (black) in regime (II) and $Ro \approx -1.82$ (red) in regime (I), where $Ro_c = -1.73$. Within the inertial wave range ($0 < \hat{\omega}\le 2$), the background turbulence level for $Ro<Ro_c$ is higher than for $Ro>Ro_c$, whereas the spectral peaks dominate for $Ro>Ro_c$. This was discussed already in section~\ref{sec:3a}. Not considered so far is the spectral tail for $\hat{\omega}>2$, outside the inertial wave range. Interestingly, beyond $\hat{\omega}=2$ (dashed line), both curves coincide well. This clear ``knee'' at $\hat{\omega}=2$ in the red curve is a feature that was observed earlier for the 3m spherical shell experiment \cite{triana_2011} confirming that those experiments were running in a more turbulent regime. This implies further that the enhanced level of turbulence for $Ro<Ro_c$ is related to inertial waves since for $\hat{\omega}>2$ no such enhancement can be seen. We think that this is another sign of a cascade to small-scale wave turbulence induced by inertial waves after the transition at $Ro_c$.

\section{Discussion and Conclusion}\label{sec:4}

In this study, we presented experimental results of a differentially rotating spherical gap flow with $0.68\cdot 10^{-5} \le E = \nu / (\Omega_o\,r_o^2) \le 1.35 \cdot 10^{-5}$ and $-2.5 < Ro = (\Omega_i-\Omega_o)/\Omega_o < 0$ (the inner sphere is rotating slower than the outer sphere). Note that these Ekman numbers are about two to three orders of magnitude larger than for most of the previous experiments \cite{kelley_et_al_2007,kelley_et_al_2010,triana_2011,rieutord_et_al_2012}. However, from a spectrogram spanned by the frequency $\hat{\omega}$ and the Rossby number $Ro$, we found excited inertial modes (approximating some of the analytical full-sphere modes found by \cite{zhang_et_al_2001}). In agreement with \cite{kelley_et_al_2007,kelley_et_al_2010,triana_2011,rieutord_et_al_2012}, almost all of our modes were non-axisymmetric and antisymmetric with respect to the equator and propagating retrograde (against the rotation of the outer shell). 

An exception was the columnar Rossby mode (5,1) excited by the Stewartson-layer instability \cite{hollerbach_2003,wicht_2014}. It was found to be persistent over the entire Rossby number range measured. In particular for $Ro>Ro_c$, our spectrograms revealed a high number of frequency couplings between the Rossby mode and the other inertial modes, leading to sidebands below and above the primary inertial mode frequency forming triads (figure~\ref{fig:6}). Similar triad interactions with Rossby modes of $m=2$ and $m=3$ have been found by \cite{hoff_et_al_2016} in a system where the inner sphere oscillates around a mean angular velocity. Therefore, we conclude that the Rossby modes in a spherical shell play a crucial role for forming triads with inertial waves.

Rieutord \textit{et al.} \cite{rieutord_et_al_2012} explained the onset of the inertial modes by the existence of \emph{critical layers} where the drift speed $\hat{\omega}/m$ of the particular inertial mode matches the angular velocity somewhere in the Stewartson layer (co-rotating resonance). The symbols in figure~\ref{fig:7} represent the radial position where the dimensional drift speed $\hat{\omega}\,\Omega_o\,r/m$ of the (5,1) mode is equal to the time-averaged azimuthal velocity. For the (3,2) and the (5,2) mode, the drift speed is by a factor of $\sim 4$ higher than the velocity inside the tangent cylinder. Therefore our measurements do not confirm the presence of critical layers for these two modes. However, as is obvious from figure~\ref{fig:7}, the (5,1) columnar Rossby mode does have a critical layer in the shear flow. For high $Ro$, close to the onset of the (5,1) mode (compare with figures~\ref{fig:3} and \ref{fig:4}), the critical layer appears first inside the tangent cylinder. With decreasing $Ro$, this layer shifts outwards because the drift speed of the mode remains constant in contrast to the mean flow.

Amazingly, the maximum Rossby number $Ro_{max}$ found by \cite{rieutord_et_al_2012} for which the (3,2) and the (5,2) mode can exist agrees surprisingly well with our data taking into account that \cite{rieutord_et_al_2012} considered the inviscid limit (see table~\ref{tab:2}). Nevertheless, our data imply that the onset of the inertial modes cannot be explained by critical layers and co-rotating resonance.

We found that after its excitation, the (3,2) mode grows significantly with decreasing $Ro$. We suggested in section~\ref{sec:3d} that the amplification can be explained by a mode-mean flow interaction that draws energy from the shear flow around the tangent cylinder when the Rossby number is decreasing. At the critical Rossby number $Ro_c$ an \emph{abrupt} transition takes place: the modes loose most of their energy and the mean flow suddenly increases. Generally speaking, the flow shows more small-scale disorder \citep{mcewan_1970,kerswell_1999} for $Ro<Ro_c$. In consequence, the inertial mode frequencies experience a shift resulting from the Doppler effect. The transition to turbulence occurs due to a secondary instability of the inertial modes. Similar instabilities could be observed in previous laboratory experiments, preferentially in precessing cavities (see e.g. \citep{mcewan_1970,kerswell_1999,lin_et_al_2015}), but also due to subharmonic instability of internal-wave attractors \cite{brouzet_et_al_2016}. It should be noted that this transition was not documented for previous spherical shell experiments with differentially rotating boundaries. The reason behind might be the particular scaling with the Ekman number. For our experiments, the occurrence of the critical Rossby number is a robust feature that scales with $E^{1/5}$. This scaling implies that previous experiments with much smaller $E$ were in regime (I) for the whole range of $Ro$ that have been considered and thus no transition was observed.

When the Ekman layers become vertical, they degenerate into another boundary layer, a process called equatorial degeneracy \cite{stewartson_1966}. The usual scaling for the layer thickness with $E^{1/2}$ breaks down and is replaced by $E^{2/5}$ with a singularity at the equator. Interestingly, the width of this blow-up equatorial region scales with $E^{1/5}$ and matches with our scaling for $Ro_c$. This suggests that for $Ro<Ro_c$ the equatorial region becomes supercritical with respect to G\"{o}rtler instability.

From \cite{koch_et_al_2013} and \cite{calkins_et_al_2010} it is known that G\"{o}rtler vortices in spherical shells with librating inner boundaries get excited in the ``blow-up" region of the equatorial boundary layer and spreading outward. Moreover, \cite{ghasemi_et_al_2016} showed that G\"{o}rtler vortices propagating from the boundary to the bulk of the fluid can drive a significant mean flow. With this knowledge, we propose the following scenario: For moderate Rossby numbers $Ro>Ro_c$, the flow is sub-critical and dominated by large-scale features. At $Ro_c$ the rotation of the inner sphere exceeds a certain critical value, the Ekman boundary layer becomes centrifugally unstable, i.e. G\"{o}rtler vortices get excited in the equatorial boundary layer where the velocity of the inner sphere is the highest. The flow becomes supercritical. Subsequently, the vortices propagate towards the vertical Stewartson layer and, according to \cite{ghasemi_et_al_2016}, amplify the mean flow around the tangent cylinder. This scenario would first explain the $E^{1/5}$ scaling of $Ro_c$ and second the massive enhancement of the zonal mean flow around the tangent cylinder at $Ro_c$ (figure~\ref{fig:7}).

We think this work is helpful for understanding, and classifying, the described flow phenomena and their instabilities in the spherical Couette flow. From a geophysical point of view, this applies not only to the dynamics of planetary interiors but also to the oceans and atmosphere, forming very thin spherical shells. From theoretical models of the equatorial dynamics, there remain large uncertainties since most of the models make use of the traditional approximation that neglects the vertical component of the Coriolis force. This might explain the disagreement between observations and theory for the ``equatorial boundary layer" as was suggested recently by Rabitti \cite{rabitti_2016}. Experiments in spherical-shell geometry are not affected by this kind of approximations and might hence form a testbed for the subtle equatorial dynamics of geophysical flows.

\section*{Acknowledgements}

	M.H. was financed by the DLR (grant no. 50 WM 0822) and is now supported by the DFG (grant no. HA 2932/7-1). We gratefully acknowledge Adrian Mazilu from the Transilvania University of Brasov (Romania) who accurately performed most of the experiments in the frame of a Traineeship ERASMUS+ program. We further thank T. Seelig for fruitful discussions and useful hints for post-processing the data. S. A. Triana thanks the European Commission for funding his study in Cottbus via the infrastructure program EuHIT. Finally, we thank C. Egbers for the opportunity to work with his spherical shell apparatus, originally funded by the DFG grant no. EG 100/1. We also thank the anonymous reviewers for their critical remarks that helped to improve the quality of the final version.




\end{document}